\documentclass[12pt,journal,epsfig,onecolumn,peerreviewca,draftclsnofoot]{IEEEtran}
\usepackage{cite}
\usepackage[dvips]{graphicx}
\usepackage{graphicx}
\usepackage{amssymb}
\usepackage{subfigure}
\usepackage{amsmath}
\newcounter{MYtempeqncnt}
\usepackage{subeqnarray}
\usepackage{cases}
\usepackage{color}
\usepackage[centerlast]{caption2}
\usepackage{wrapfig}
\usepackage{url}

\usepackage{siunitx} 
\usepackage{multirow} 
\usepackage{booktabs} 
\usepackage{longtable} 
\usepackage{rotating} 

\usepackage{algorithm}
\usepackage[noend]{algpseudocode}

\begin{document}

\title{Optimal Pricing for Job Offloading in the MEC System with Two Priority Classes}

\author{Lingxiang~Li, Marie Siew,
Tony~Q.S.~Quek, ~\IEEEmembership{Fellow,~IEEE,}
Zhi~Chen, ~\IEEEmembership{Senior Member,~IEEE}
\IEEEcompsocitemizethanks{
\IEEEcompsocthanksitem Lingxiang Li and Zhi Chen are with the National Key Laboratory of Science and Technology on Communications,
University of Electronic Science and Technology of China, Chengdu 611731. 
E-mail: \{lingxiang.li, chenzhi\}@uestc.edu.cn
\IEEEcompsocthanksitem Marie Siew and Tony Q.S. Quek are with Information Systems Technology and Design Pillar,
Singapore University of Technology and Design, Singapore 487372. 
E-mail: \{marie\_siew@mymail.sutd.edu.sg, tonyquek@sutd.edu.sg\}}
}

\maketitle


\begin{abstract}
Multi-Access edge computing (MEC) is an emerging paradigm where users offload
computationally intensive jobs to the Access Point (AP).
Given that the AP's resources are shared by selfish users, pricing is a useful
tool for incentivising users to internalize the negative externality of
delay they cause to other users.
Nevertheless, different users have different negative valuations towards delay as some are more delay
sensitive. To serve heterogeneous users, we propose a priority
pricing scheme where users can get served first for a higher price.
Our goal is to find the prices such that in decision making,
users will choose the class and the offloading frequency that jointly maximize social welfare.
With the assumption that the AP knows users' profit functions,
we derive in semi-closed form the optimal prices.
However in practice, the reporting of users's profit information
incurs a large signalling overhead. Besides, in reality users
might falsely report their private profit information.
To overcome this, we further propose a learning-based pricing mechanism where no knowledge of individual
user profit functions is required.
At equilibrium, the optimal prices and average edge delays are learnt,
and users have chosen the correct priority class and offload at the
socially optimal frequency.
\end{abstract}

\begin{IEEEkeywords}
Multi-access edge computing, differentiated services, priority pricing, decentralized mechanism.
\end{IEEEkeywords}

{\section{Introduction}}
With the recent advancements of technology, end devices now run novel and diverse applications such as mobile gaming, facial recognition, and augmented reality. These applications are computationally intensive and energy consuming.
The computational capabilities and battery power of end devices, although improving over the past few years, are still relatively insufficient in meeting these requirements \cite{Ericsson15}.
To deal with this insufficiency, computationally intensive jobs are typically offloaded to cloud servers via access points (e.g., base stations or WiFi access points), helping to prolong battery life and improve user experience.
However, jobs involving augmented reality, human computer interaction also carry stringent low latency requirements.
With cloud computing, it is difficult to meet these requirements, as the huge distance between the cloud and the users results in users experiencing wide area network (WAN) delay \cite{MSat09}.
In this regard, the trend of edge computing is occurring \cite{Pavel17,YuyiMao17,Wei18}, wherein job computation or data storage will be shifted to the network edge, instead of faraway clouds. In this way, WAN delay is avoided.
This kind of computing refers to Multi-Access Edge Computing (MEC), wherein
the network edge refers to the access points (APs).

Since the AP is also in control of the wireless channels,
there is substantial interest in investigating the joint
optimization of radio and computing resource allocation in MEC \cite{Changsheng16,Yuyi17,Changsheng17,Chen17,Fei18,Ying2019}. This enhances the system wide resource allocation efficiency.
Besides, as compared to the cloud, the network edge usually has limited computational
resources and tasks might need to wait in a queue before execution. Thus,
the works \cite{Jeongho15,Guanglin18,Ouyang18,ChenFeng17,Seung18,Shuaishuai18} further take queueing delay into consideration. Specifically,
\cite{Jeongho15,Guanglin18,Ouyang18,ChenFeng17} proposed
dynamic offloading algorithms based on Lyapunov optimization,
incorporating the long-term average cost minimization problem into real-time optimization.
The works \cite{Jeongho15,Guanglin18} study the single user scenario with a user running various types of tasks;
the works \cite{Ouyang18,ChenFeng17} further investigate the multiple users scenario with users moving erratically
or under a probabilistic constraint on the queue length violations.
On the other hand, the works \cite{Seung18,Shuaishuai18} take the
average queueing delay as performance metric.
Specifically, the work \cite{Seung18} applies stochastic geometry and parallel computing to derive
the scaling laws of communication/computation latency with respect to network-load parameters.
The work \cite{Shuaishuai18} considers concurrent traditional uplink/downlink transmissions and MEC-based
tasks, differentiating tasks via quality of service (QoS) requirements and
assigning higher priorities to those tasks with lower latency requirements.

To the best of the authors' knowledge, few works
use priority (in a queue) as a mechanism to optimally allocate radio and compute resources in MEC.
Priority in a queue is an important consideration in optimizing resource
scheduling, since users and jobs are heterogenous, and have different costs or negative valuations
towards waiting time \cite{SRao98}. For example, an application involving
facial recognition or mobile gaming is
more delay sensitive compared to
an internet browsing application, as the first two involve
interaction and real time feedback \cite{Jeongho15,ChenFeng17}. Providing a
priority option allows users and jobs with higher waiting costs
to get served first. In contrast,
providing only one service class will incur welfare loss for
users with higher waiting costs. Many service industries for
example telcos, the postal service and theme parks all offer a
priority option. It increases the service
provider's revenue and allocates resources in a
way to maximize the overall welfare of heterogeneous users.
With this in mind, in this paper we will treat priority as a valuable resource and approach resource optimization in MEC via a priority mechanism.


\subsection{Related works and main contributions of this paper}
A large body of current works formulate the joint
optimization of radio and computing resource in MEC as a centralized
optimization problem, in particular as a Mixed Integer Nonlinear Programming (MINLP)
problem \cite{Changsheng16,Yuyi17,Changsheng17,Chen17,Fei18,Ying2019}.
However, for centralized optimization to be implemented, the AP must be able to control users' offloading amount.
Secondly, the AP must have knowledge of users' offloading profit functions.
In this work, we aim to optimize resource allocation for MEC in a distributed manner, given that centralized optimization requires control and perfect information which the AP may not have.
Other works taking this distributed approach include the following: in \cite{Lyu17} the initial centralized MINLP problem is broken down into sub-problems and dealt with semi-distributively.
\cite{Xuchen15,Xuchen16,Guo16,Jianchao18,Ling16} reformulate the centralized problem into a game
and derive the Nash Equilibrium.
This equilibrium suffers from welfare loss because in decision making,
selfish users did not internalize the negative
externalities of congestion and delay they cause to the system.
In light of this and because the AP cannot directly control users' decisions, we propose using pricing to make users internalize the negative externality in decision making.

Pricing is a key tool to allocate resources to the users who value them the most, thus controlling congestion in resource competition cases \cite{Courcoubetis03}.
However, only a small body of works have studied MEC from an economics viewpoint.
Existing works from the economics viewpoint such as \cite{Shih2019,Weiwei18,Abbas17} are based on abstract utility functions (e.g., a simple logarithmic function), and do not
capture various user-side factors, such as the processing capacity
of end devices and the amount of remaining computing tasks. The work \cite{Yeongjin18}
takes those user-side factors into account, and proposes Lyapunov-based algorithms
for energy/monetary cost minimization while ensuring finite processing delay.
Similar to other Lyapunov-based algorithms proposed by \cite{Jeongho15,Guanglin18,Ouyang18,ChenFeng17},
the implementation of this work requires sharing information such as
the queueing states of users and the AP at each time slot, which the AP has to periodically measure.
As such, implementing these algorithms will incur a large signalling overhead.

In \cite{Lingxiang19TMC}, we have considered the homogeneous user scenario and proposed an incentive-aware job offloading control framework, by introducing an economics model encapsulating
physical layer factors, such as the wireless channel condition,
the computational energy, the computational and queueing delay of both local and edge computing.
The average delay (via statistical measurements over time) is considered
since the prices should not fluctuate too often. 
We have shown that the proposed framework can help regulate users'
behaviors via pricing efficiently \cite{Lingxiang19TMC}. Further, in \cite{Lingxiang_ICASSP2019},
we have proposed a learning based pricing mechanism for the homogeneous user scenario, in which with no direct
control and no knowledge of users' private information such as queueing states, the AP
learns the optimal price, inducing self-interested users to make socially optimal offloading decisions.
In this work, we aim to extend our previous results to the heterogeneous user case
and provide the option of priority, where the MEC server serves the high priority users first for a higher service fee.


Our main contributions are summarized below.
\begin{enumerate}
\item
We introduce a novel utility function (see Eq. (\ref{eq13})), which, after subtracting the delay cost
required by edge computing, equals the profit a user achieves by offloading (see Eq. (\ref{eq12})).
This utility function measures 
the difference between the local computing cost and the offloading cost (including the offloading delay and the offloading energy consumption).
Besides, this utility function satisfies the basic nature of utility functions in economics, thus
bridging the gap between the economic and physical layer parameter optimizations.
\item \emph{Complete users' individual demand information:}
Based on the given utility function we propose an incentive-aware job offloading framework, in which
the user decides the amount to offload by comparing its utility function with the delay cost of edge computing. We characterize the socially optimal point, as a function of the users' profit functions as well as the average edge computing delays (see Theorem \ref{theorem1}).
Furthermore, we derive in semi-closed form the prices which can incentivise users to choose the \emph{correct} priority class and offload at a socially optimal frequency.
\item \emph{Without users' individual demand information:}
We propose a learning-based pricing mechanism where no knowledge of individual user profit functions is required (see {Algorithm} \ref{alg:NE2}). Instead, based on the measured congestion levels, the AP broadcasts prices and expected edge delays of the distinct priority classes. At equilibrium, the optimal prices and expected edge delays are learnt, with which the AP induces self-interested users to choose the \emph{correct} priority class and make socially optimal offloading decisions, thus maximizing the system-wide welfare in a decentralized way (see Theorem \ref{theorem2} and \ref{theorem3}).
\end{enumerate}

\newtheorem{proposition}{Proposition}
\newtheorem{theorem}{Theorem}
\newtheorem{corollary}{Corollary}
\newtheorem{lemma}{Lemma}

\section{Communication/Edge Computing System Model}

We consider an MEC system consisting of an access point (AP) and $N$ end users.
The wireless AP could be a base station, or a Wi-Fi access point.
Other than being a conventional access point to the core network, it is installed with an additional edge computing server.
The end devices might be running computation-intensive and delay-sensitive jobs, and may have insufficient computing power or limited battery energy to complete those jobs.
As such, they may offload part/all of their jobs to the AP.
In this section, we will introduce the offloading policy, the wireless channel model,
followed by the models for computing in detail (see Fig. 1).

\begin{figure}[!h]
\centering
\includegraphics[width=3in]{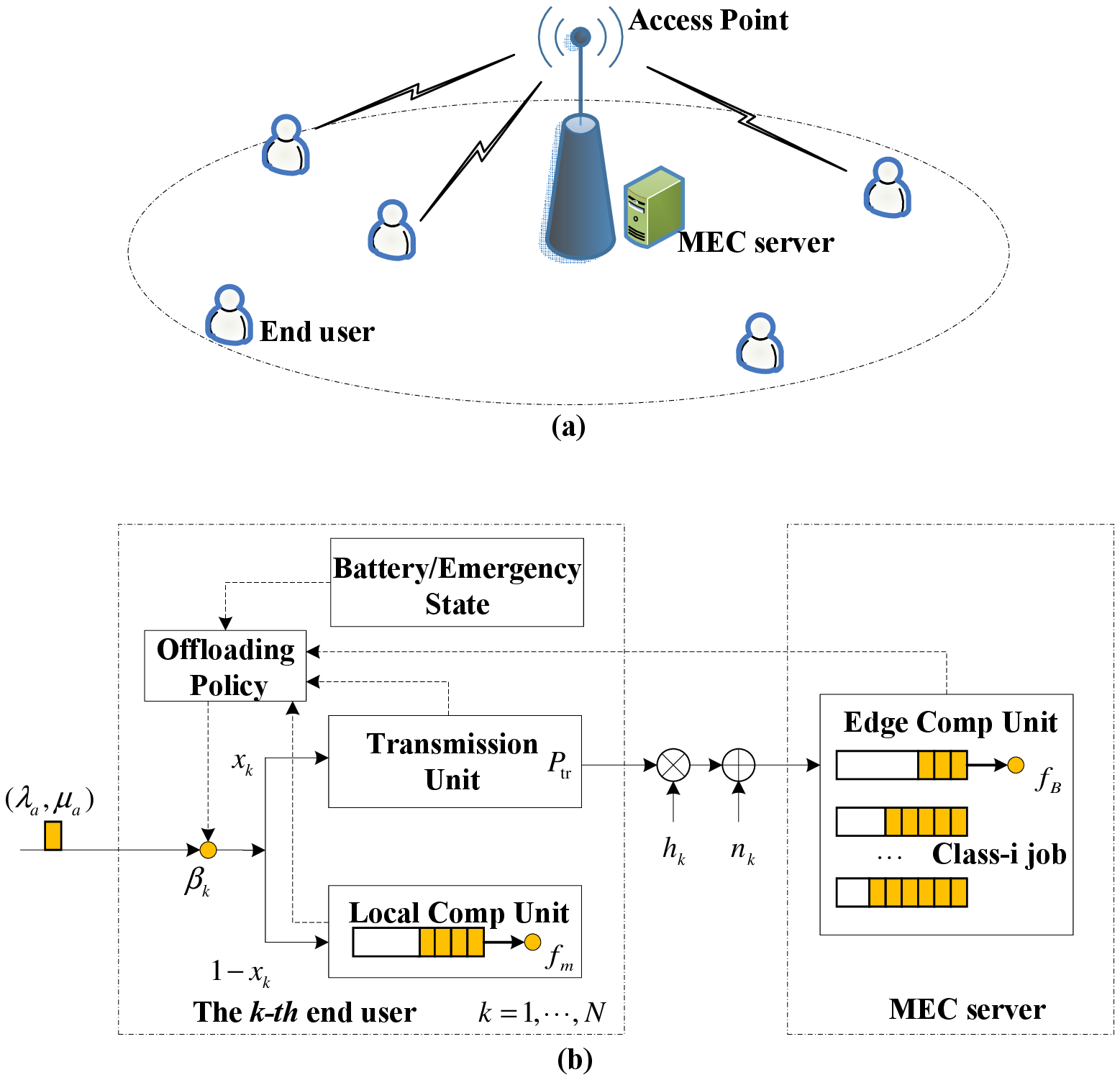}
\DeclareGraphicsExtensions. \caption{An MEC system illustration with job arrival, offloading and computation.}
\label{fig1}
\vspace* {-12pt}
\end{figure}

\subsection{Job generation model and offloading policy}
Jobs arrive at the end users following a Poisson process of rate $\lambda_a$.
The service time of a job is identically distributed (i.i.d.) and exponentially distributed,
with an average of $L_a$ bytes input data to offload.
Denote the processing density by $B_a$.
Then the average CPU cycles to run for a job equals $\mu_a=L_aB_a$. 

This paper considers the scenario with flat-fading channels,
and assume that the user can finish offloading in a channel block.
Hence, we consider the following \textit{offloading policy}.
When a job arrives, if the achievable rate is higher than
a threshold $\beta_k$, the end user would
offload its job to the AP; otherwise, the end user will choose local computing.

\subsection{Radio access model}
Users \textit{access the AP in an FDMA mode},
suffering no multi-user interference.
Let $h_k$ denote the small-scale channel gain from user $k$ to the AP.
The achievable uplink data rate is thus,
\begin{align}
R_k = {\rm log}(1+d_k^{-\alpha}|h_k|^2P_{\rm tr}/\sigma^2),\ k=1,2,\cdots, N,\label{eq1}
\end{align}
where $d_k$ denotes user $k$'s distance to the AP and $\alpha$ represents the path loss exponent.
$P_{\rm tr}$ is the transmission power and $\sigma^2$ denotes the received noise power at the AP.
In addition, by comparing the achievable data rate $R_k$ with the expected data rate $\beta_k$ and by Shannon's theorem \cite{David05}:
when $R_k>\beta_k$ the end device can transmit
its job to the AP successfully. Hence, the user $k$'s offloading frequency (probability) is
\begin{align}
x_k={\rm Pr}(|h_k|^2 > (e^{\beta_k}-1)\rho_k^{-1}), \  k=1,2,\cdots, N, \label{eq2}
\end{align}
where $\rho_k \triangleq d_k^{-\alpha}P_{\rm tr}/\sigma^2$.
A user could control its offloading frequency by adjusting the threshold $\beta_k$.

\subsection{Computation model}
Based on the above radio access model and offloading policy,
we discuss the total overhead/cost of local computing and edge computing.
Both the processing delay and buffer delay are taken into consideration.

\subsubsection{Local computing}
By the Poisson arrival process and exponential job service time assumptions,
we have an M/M/1 queue for local computing.
Let $f_m$ be the end device's computing capability (CPU cycles
per second), then the expected time spent per job (including both the job execution time
and the time spent awaiting in a local buffer)
is
\begin{align}
D_k^{\rm LC}(x_k)&={1}/({\mu_m-\lambda_a\bar{x}_k}), \quad  k=1,2,\cdots, N.\label{eq3}
\end{align}
where the local computing probability $\bar{x}_k \triangleq1-x_k$, and the service rate $\mu_m\triangleq f_m/\mu_a$.
Next, the computational energy of local computing is
\begin{align}
E_k^{\rm LC}&=\kappa_m f_m^2 \mu_a, \  k=1,2,\cdots, N, \label{eq4}
\end{align}
where $\kappa_m f_m^2$ is the power consumption the end user runs one CPU cycle,
and $\kappa_m$ is an energy consumption coefficient that depends on the chip architecture \cite{Antti10}.

The total weighted cost of local computing is \cite{Xuchen15,Xuchen16}
\begin{align}
Z_k^{\rm LC}(x_k)&=c_k^e E_k^{\rm LC}+c_k^d D_k^{\rm LC}(x_k), \  k=1,2,\cdots, N,\label{eq5}
\end{align}
where $0<c_k^e<1$ (in units 1/Joule) and $0<c_k^d<1$ (in units 1/Second) are the weights of
computational energy and delay.
The weights allow different users to
place different emphasis in decision making.
For example, if the end device is at a low battery state,
it would give energy consumption more emphasis,
choosing a bigger value of $c_k^e$.
If the user is running urgent jobs, it would give the delay cost more emphasis.
In this paper, we consider two classes of resource-hungry users with different service requirements,
i.e., one class of users setting $c_k^d=c_{\rm H}^d$, $k=1,\cdots,N_{\rm H}$,
and the other setting $c_k^d=c_{\rm L}^d$, $k=N_{\rm H}+1,\cdots,N_{\rm H}+N_{\rm L}$,
where $N_{\rm H}+N_{\rm L}=N$.

\begin{table}[t!]
  \begin{center}
    \caption{Summary of key notations}
    \label{tab:table1}
    \begin{tabular}{l|S}
      \toprule
      \textbf{Notation} & \textbf{Description} \\
      \hline
      $\lambda_a$ & \textrm{Job arrival rate at the end device (jobs per second)} \\
      \hline
      $\mu_a$ & \textrm{Average CPU cycles needed by the computation job} \\
      \hline
      $B_a$ & \textrm{Processing density (in cycles/bit)} \\
      \hline
      $L_a$ & \textrm{The input data size (in Bytes)} \\
      \hline
      $\kappa_m$ & \textrm{The energy coefficient of the end device} \\
      \hline
      $x_k$ & \textrm{User $k$'s offloading frequency} \\
      \hline
      $d_k$ & \textrm{User $k$'s distance to the access point} \\
      \hline
      $P_{\rm tr}$ & \textrm{Each user's transmit power} \\
      \hline
      $\sigma^2$ & \textrm{Received noise power at the access point} \\
      \hline
      $c_k^e$ & \textrm{User $k$'s weight of the computational energy} \\
      \hline
      $c_k^d$ & \textrm{User $k$'s weight of the computational time} \\
      \hline
      $h_k$ & \textrm{User $k$'s instantaneous small-scale channel gain} \\
      \hline
      $\beta_k$ & \textrm{User $k$'s transmission rate} \\
      \hline
      $f_m$ & \textrm{CPU-cycle frequency of the end device} \\
      \hline
      $\mu_m$ & \textrm{Computing service rate of User $k$ (jobs per second)} \\
      \hline
      $f_B$ & \textrm{CPU-cycle frequency of the MEC server} \\
       \hline
      $\mu_B$ & \textrm{Computing service rate of the MEC server (jobs per second)} \\
       \hline
      $g_k$ & \textrm{User $k$'s demand function} \\
      \hline
      $N$ & \textrm{The total number of end users} \\
      \bottomrule
    \end{tabular}
  \end{center}
\end{table}
\subsubsection{Edge computing}
First, the time taken to offload to the AP is
\begin{align}
D_{k,1}^{\rm EC}(x_k)&={l_a\mu_a}/{\beta_k (x_k)}, \  k=1,2,\cdots, N.\label{eq6}
\end{align}
This indicates that the energy required by offloading is
\begin{align}
E_k^{\rm EC}(x_k)&=P_{\rm tr}{l_a\mu_a}/{\beta_k (x_k)}, \  k=1,2,\cdots, N.\label{eq7}
\end{align}

Subsequently, the offloaded job will stay at the AP's buffer until it leaves after execution.
For the purpose of increasing the net welfare, the AP applies priority queueing,
where it provides different qualities of service to different
classes of traffic. Suppose that traffic classes are partitioned into two sets, ${\rm H}$ and $\rm L$.
The job execution is FCFS, except that the jobs from the class ${\rm H}$ are always given priority over those from the class ${\rm L}$, and the execution of an ordinary user's job is preempted if a priority user's job arrives.

Based on the aforementioned offloading policy, the end device will choose to offload its jobs with probability $x_k$.
This, combined with the \textit{splitting} property of the queueing theory \cite{Refael03},
indicates that the job offloading from a end user also follows a Poisson process, with parameter $\lambda_a x_k$.
Hence, the job arrival at the AP is a superposition of multiple Poisson processes from multiple end users, which, according to the \textit{superposition} property of queueing theory \cite{Refael03}, is another Poisson process with arrival rate as the sum arrival rate of the superposed processes.
Therefore, the arrival rate at the AP is $\sum\nolimits_{{\rm i}_j={\rm H}} \lambda_a x_j$ for the higher priority class of users and $\sum\nolimits_{{\rm i}_j={\rm L}} \lambda_a x_j$ for the lower priority class of users.
On the other hand, the service times are i.i.d. and exponentially distributed with parameter $\mu_a$.
As such, we get an M/M/1 queue with two priority classes for computing at the AP.

Let $f_B$ be the AP's computing capability (CPU cycles
per second). The service rate of the end device is then $\mu_B=f_B/\mu_a$ (jobs per second).
Thus, for the user choosing the first priority, the expected time a job spending at the AP
(including both the computation execution time and the time spent awaiting in the edge buffer)
can be expressed as,
\begin{align}
D_{{\rm H}}^{\rm EC}({\bf x})&={1}/({\mu_B-\sum\nolimits_{{\rm i}_j={\rm H}} \lambda_a x_j}),\label{eq8} 
\end{align}
where ${\bf x} \triangleq(x_1, x_2, \cdots, x_N)$. 
For the user choosing second priority, the expected time a job spends at the AP
can be expressed as,
\begin{align}
D_{{\rm L}}^{\rm EC}({\bf x})&=\dfrac{\mu_B D_{\rm H}^{\rm EC}({\bf x})}{\mu_B-\sum\nolimits_{{\rm i}_j={\rm L}} \lambda_a x_j-
\sum\nolimits_{{\rm i}_j={\rm H}} \lambda_a x_j}.\label{eq9}
\end{align}

We neglect the energy overhead of edge computing as \cite{Xuchen16,Fei18,Jianchao18}, since
normally the AP can access to wired charing and it has no lack-of-energy issues.
Also, similar to many other related works such as \cite{Xuchen15,Xuchen16,Guo16,Jianchao18,Ling16,Lyu17},
we neglect the time overhead for the edge to send the computation outcome back to the end user, since for many applications (e.g., facial recognition) the
size of the computation outcome in general is much smaller
than the size of computation input data, including the device
system settings, program codes and input parameters. As such, we assume that the result of an offloaded task is guaranteed on the downlink direction.
Combining (\ref{eq6}) to (\ref{eq9}) yields the total weighted cost of edge computing by user $k$, i.e.,
\begin{align}
Z_k^{{\rm{EC}}}({\rm i}_k, {\bf{x}})
= c_k^eE_k^{{\rm{EC}}}({x_k}) + c_k^d(D_{k,1}^{{\rm{EC}}}({x_k}) + D_{{\rm i}_k}^{{\rm{EC}}}({\bf{x}})).\label{eq10}
\end{align}
Note that the total cost by offloading, as well as the edge computing buffering delay
$D_{{\rm i}_k}^{{\rm{EC}}}({\bf{x}})$,
depends on both the local variable (i.e., $x_k$ for the $k$-th user)
and also the offloading frequency of other users.
As we will see later, due to this coupled nature of delay we have coupled objective functions.

\section{Proposed Economics Model for Multi-Access Edge Computing and Problem Formulations}
By the aforementioned offloading policy, when a job arrives the user will
offload with probability $x_k$, 
and locally compute with probability $\bar{x}_k$.
Hence, the expected total cost of offloading is
\begin{align}
Z_k({\rm i}_k, {\bf x}) &= \bar{x}_k Z_k^{\rm LC}({x_k}) + x_k  Z_k^{\rm EC}({\rm i}_k, {\bf x}). \label{eq11}
\end{align}

On the other hand, when there is no such edge server providing computing power, users have
to run jobs locally with average cost
$Z_k^{\rm LC}(0)$.
Therefore the gross profit of
offloading by the $k$-th user under a given offloading strategy ${\bf x}$ is,
\begin{align}
V_k({\rm i}_k, {\bf x}) &=Z_k^{\rm LC}(0)-Z_k({\rm i}_k, {\bf x}), \  k=1, \cdots, N. \label{eq12}
\end{align}

The key idea of the economics model is to introduce the utility function and the cost function.
Some observations are in order. Firstly, the profit each user obtains equals the cost savings
from offloading, and it is a linear combination of the
energy costs and the delay costs.
Secondly, the coupled delay cost $D_{{\rm i}_k}^{\rm EC}({\bf x})$ reflects the harm/congestion each user
causes to the other users, with a bigger value indicating that the AP provides worse service to the users.
Thirdly, except for the expected time spent at the AP, i.e., $D_{{\rm i}_k}^{\rm EC}({\bf x})$, which depends
on all users' offloading decisions, the other items in the profit function
only depend on each user's own offloading frequency $x_k$.
Motivated by these observations, we introduce a utility function $U_k(x_k)$ which includes the
items in the profit function that only depend on the local variable $x_k$, i.e.,
\begin{align}
U_k(x_k)=&Z_k^{\rm LC}(0)-\bar{x}_k Z_k^{\rm LC}({x_k}) \nonumber \\
&- x_k(c_k^e E_k^{\rm EC}(x_k)+c_k^d D_{k,1}^{\rm EC}(x_k)).  \label{eq13}
\end{align}
This, combined with (\ref{eq12}), indicates that
\begin{align}
V_k({\rm i}_k, {\bf x})=U_k(x_k)-C_k({\rm i}_k,{\bf x}), \  k=1, \cdots, N,   \label{eq14}
\end{align}
where $C_k({\rm i}_k,{\bf x}) \triangleq c_k^d x_k D_{{\rm i}_k}^{\rm EC}({\bf x})$ can be regarded as the delay cost caused by sharing an MEC server at the AP.

Taking the derivative of $U_k(x_k)$ with respect to $x_k$, we arrive at
the demand function of user $k$, i.e.,
\begin{align}
&g_k({x}_{k})\triangleq  {\partial U_k(x_k)}/{\partial x_k}. \nonumber
\end{align}
For each end user, it holds true that the
demand function $g_k({x}_{k})$ is a monotonically decreasing function of the offloading frequency $x_k$.
Moreover, there exists a unique solution of the equation $g_k({x}_{k})=0$, denoted as $x_k^{up}$.

\begin{figure}[!h]
\centering
\includegraphics[width=3in]{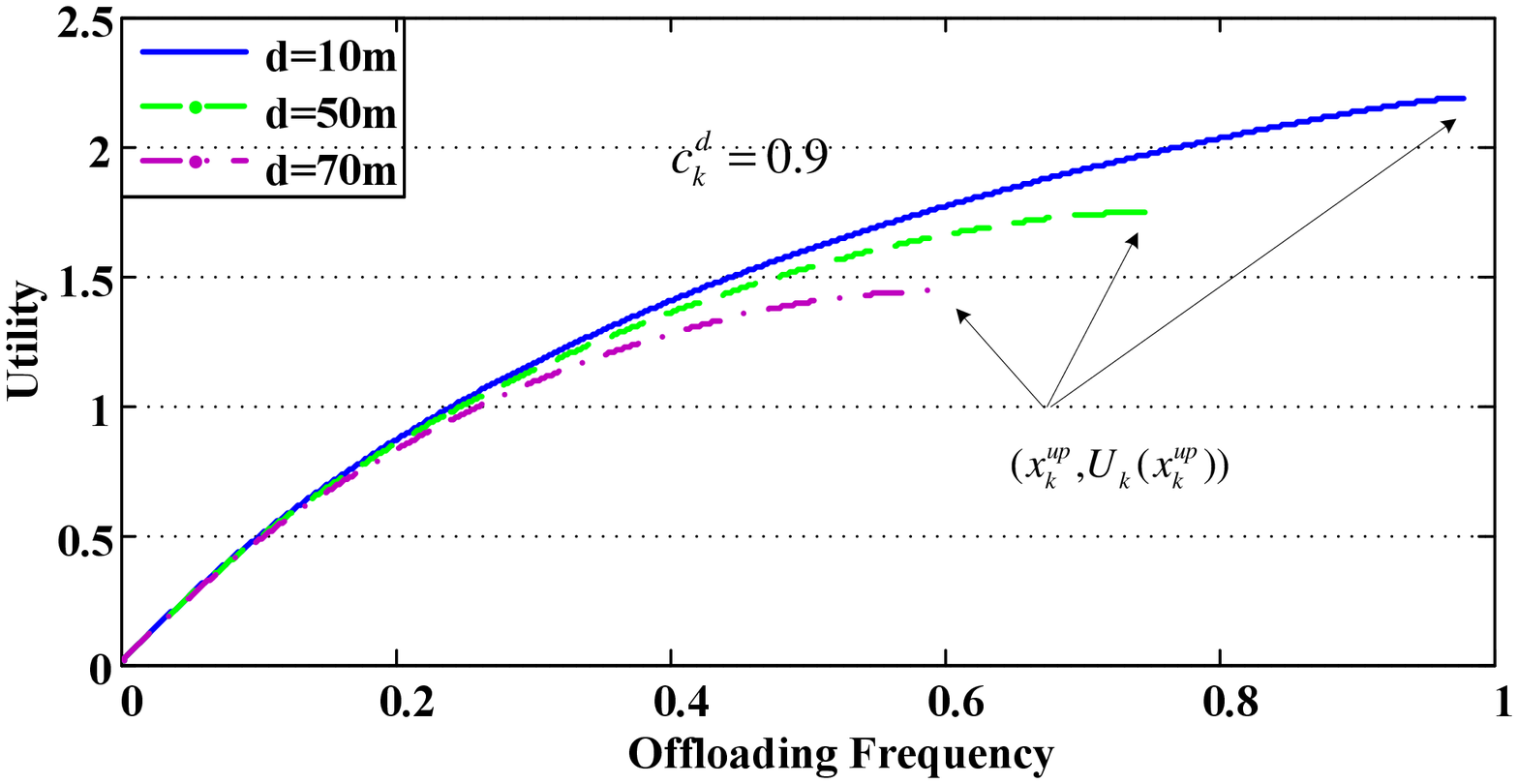}
\DeclareGraphicsExtensions. \caption{The achievavble utility for varying distances to AP.}
\label{fig0}
\end{figure}

The utility function in (\ref{eq13}) measures the welfare of offloading, while containing physical layer meaning.
It is illustrated in Fig. \ref{fig0}, for
a system with parameters set as in the numerical results section.
Different distances, i.e., $d=10$m, $d=50$m and $d=70$m, from the user to the AP are considered.
One can see that the utility function is strictly increasing while the rate of increase, i.e., the demand,
decreases as $x_k$ increases; this is consistent with the law of diminishing marginal returns, a typical property of  utility functions in economics.
Moreover, the user closer to the AP has more offloading demand and can achieve a higher utility for the same offloading amount.
This agrees with intuition that a user nearer the AP experiences a better wireless channel.

\subsection{Problem formulations}

We consider the social-welfare offloading decision problem, where the objective is to maximize the net social welfare $\sum\nolimits_{k=1}^{ N} V_k({\bf x})$ as follows.

\underline{\textbf{Social Problem:}}
\begin{align}
\{\bar{\rm i}_k^\star, {\bar x}_k^{\star}\} \triangleq &{\arg}  \max_{\{{\rm i}_k, x_k\}} \quad \sum\nolimits_j V_j({\rm i}_j,{\bf x}), \nonumber \\
& {\rm s.t.} \quad  {\rm i}_k\in \{{\rm H, L}\}, 0 \le x_k  \le 1, \forall k. \label{eq15}
\end{align}


Generally, the above Social Problem can be solved centrally, which requires control and perfect utility information of users.
As an alternative, in this work we aim to provide users with the appropriate incentives to control their flows and maximize the overall system performance in a distributed way.

Intuitively, to maximize the net social welfare each user should also be concerned with the congestion it causes to other users and should keep its offloading under an appropriate amount for other users' welfare; the difficulty lies in how to incentivise users to do so when
users are selfish and will choose their offloading decisions such that their
individual profit $V_k(x_k)$ is maximized.

Pricing is a useful tool in incentivising users to choose the socially optimal levels of demand.
The key idea is to enforce users to pay for the congestion they cause to the other users.
In the following Regulated Selfish Problem, we study the pricing-based scheme, which
charges users an additional edge computing service fee to regulate users' behavior.

\underline{\textbf{Regulated Selfish Problem:}}
\begin{align}
\{{\rm i}_k^\star, { x}_k^{\star}\} \triangleq &{\arg}  \max_{\{{\rm i}_k, x_k\}} \quad U_k(x_k)-(c_k^d D_{{\rm i}_k}^{\rm EC}+p_{{\rm i}_k})x_k, \nonumber \\
& {\rm s.t.} \quad  {\rm i}_k\in \{{\rm H, L}\}, 0 \le x_k  \le 1, \forall k. \label{eq16}
\end{align}
where $p_{\rm H}$ and $p_{\rm L}$ respectively denote the unit service fee for the MEC service with high priority and low priority.

Does there exist price pair $(p_{\rm H},p_{\rm L})$ such that the individual objectives of self-interested users will be aligned with the social welfare maximization objective?
If it exists, how to design pricing schemes to achieve the highest net social welfare?
In the following sections, we answer the above two questions.

\section{The Socially Optimal Offloading Mechanism with Two Priority Classes of Users}
In this section, we study the socially optimal offloading mechanism in a complex scenario with two priority classes of users.
The priority queue is offered, wherein other than optimizing over each user's frequency of offloading, we also
need to optimize over the job scheduling sequence.
This is because as mentioned earlier, different users might have different negative valuations towards delay experienced. Therefore, providing one class of service will incur further welfare loss, on the part of users
have high delay sensitivity. To deal with heterogeneous users, our mechanism provides the option of priority, where it serves the high priority users first for a higher service fee.
Distinct prices for the two priority classes will ensure truthfulness, preventing all users from declaring priority. The prices will be set such that in decision making, when users weigh their payoffs from the two classes, they will choose the class that maximizes social welfare.

In what follows, we first analyze the structural property of the formulated social and regulated selfish problems
and admit in semi-closed form the optimal prices that induce socially optimal offloading decisions, which, however,
requires users' individual profit functions.
As such, in the second subsection, we propose
an evolutional pricing algorithm, which requires no individual utility functions
and learns the \textit{correct} prices based on the congestion level.

\subsection{Optimal price for offloading under complete information}
Before proceeding, we first introduce a lemma as follows.
\begin{lemma}\label{lemma1}
\textit{For the purpose of minimizing the total delay cost,
the end users putting more consideration to the delay cost, i.e.,
$c_k^d=c_{\rm H}^d$, shall be given the higher priority service, i.e., ${\rm i}_k={\rm H}$.
}
\end{lemma}
\begin{IEEEproof}
According to the famous $c\mu$ \textit{rule} (or, here, the $c_k^d\mu_a$ \textit{rule}) \cite{Beja75},
providing a higher priority to those users with a higher value of delay cost, is optimal in terms of minimizing the system-wide delay cost.
This completes the proof.
\end{IEEEproof}
\medskip

Based on Lemma \ref{lemma1}, the users with higher delay costs
shall be given a higher priority service, while those with lower delay costs
shall wait in the queue for the AP finish executing higher priority jobs.
Therefore, the expression of delays in (\ref{eq8}) and (\ref{eq9}) can be
respectively rewritten as,
\begin{subequations}
\begin{align}
D_{{\rm H}}^{\rm EC}({\bf x})&={1}/{\Psi_{\rm H}}, \label{eq26a}\\
D_{{\rm L}}^{\rm EC}({\bf x})&={\mu_B D_{\rm H}^{\rm EC}({\bf x})}/{\Psi}. \label{eq26b}
\end{align}
\end{subequations}
where $\Psi_{\rm H}\triangleq {\mu _B} - \sum\nolimits_{c_j^d = c_{\rm{H}}^d} {{\lambda _a}} {x_j}$ and
$\Psi\triangleq {\mu _B} - \sum\nolimits_{j} {{\lambda _a}} {x_j}$.

For the Social Problem, by the first order condition, at the maximum is holds that
\begin{align}
&\dfrac{\partial U_k(x_k)}{\partial x_k}- \sum\nolimits_{j=1}^{ N}\dfrac{\partial c_j^d x_j D_{{\rm i}_j}^{\rm EC}({\bf x})}{\partial x_k}=0 \nonumber \\
\Leftrightarrow& g_k(x_k)= c_k^d D_{{\rm i}_k}^{\rm EC}({\bf x})+\sum\nolimits_{j=1}^{ N}\dfrac{\partial c_j^d  D_{{\rm i}_j}^{\rm EC}({\bf x})}{\partial x_k}x_j. \label{eq21}
\end{align}

Substituting (\ref{eq26a}) and (\ref{eq26b}) into (\ref{eq21}) and
by further derivations, we arrive at the following theorem.
\medskip
\begin{theorem}\label{theorem1}
\textit{Considering an MEC network consisting of two priority classes of users,
the optimal offloading frequencies of the end users are given by solving the
following equations,
\begin{align}
g_k(x_k)&= c_k^d D_{{\rm i}_k}^{\rm EC}({\bf x})+\sum\nolimits_{j=1}^{ N_{\rm H}}\frac{\partial c_{\rm H}^d  D_{\rm H}^{\rm EC}({\bf x})}{\partial x_k}x_j \nonumber \\
&+\sum\nolimits_{j=1}^{ N_{\rm L}}\frac{\partial c_{\rm L}^d   D_{\rm L}^{\rm EC}({\bf x})}{\partial x_k}x_j, \ k=1,2,\cdots, N, \label{eq22}
\end{align}
which equals solving the equations in (\ref{eq23}) on the top of next page.
}
\end{theorem}


\begin{figure*}[!t]
\normalsize
\setcounter{MYtempeqncnt}{\value{equation}}
\setcounter{equation}{19}
\begin{align}
{g_k}({x_k}) = \left\{ {\begin{array}{*{20}{c}}
{c_k^dD_{\rm{H}}^{{\rm{EC}}}({\bf{x}}) + \sum\nolimits_{j = 1}^{{N_{\rm{H}}}} {\dfrac{{{\lambda _a}c_{\rm{H}}^d{x_j}}}{{\Psi _{\rm{H}}^2}}}  + \sum\nolimits_{j = 1}^{{N_{\rm{L}}}} {\dfrac{{{\lambda _a\mu _B}({\Psi _{\rm{H}}} + \Psi )c_{\rm{L}}^d{x_j}}}{{{\Psi ^2}\Psi _{\rm{H}}^2}}} ,}&{{\rm{if}}\;c_k^d = c_{\rm{H}}^d,}\\
{c_k^dD_{\rm{L}}^{{\rm{EC}}}({\bf{x}}) + \sum\nolimits_{j = 1}^{{N_{\rm{L}}}} {\dfrac{{{\lambda _a}{\mu _B}c_{\rm{L}}^d{x_j}}}{{{\Psi ^2}{\Psi _{\rm{H}}}}}} ,}&{{\rm{if}}\;c_k^d = c_{\rm{L}}^d.}
\end{array}} \right.\label{eq23}
\end{align}
\setcounter{equation}{\value{MYtempeqncnt}}
\vspace*{-4pt}
\end{figure*}
\addtocounter{equation}{1}

\begin{IEEEproof}
See Appendix A.
\end{IEEEproof}
\medskip

\emph{Remark 1:} By solving the $N$ equations in (\ref{eq22}), we obtain
the offloading solutions of the Social Problem. Furthermore,
this solution is unique.
\medskip

In contrast, for the Regulated Selfish Problem, at the maximum it holds that
\begin{align}
g_k(x_{k})=c_k^d D_{{\rm i}_k}^{\rm EC}+p_{{\rm i}_k},\ \forall k.  \label{eq24}
\end{align}
Comparing (\ref{eq24}) with (\ref{eq22}) and for the purpose of
aligning the individual objectives with
the social welfare objective, the price shall be set as,
\begin{align}
p_{{\rm i}_k}=\sum\nolimits_{j=1}^{ N_{\rm H}}\frac{\partial c_{\rm H}^d  D_{\rm H}^{\rm EC}({\bf x})}{\partial x_k}x_j +\sum\nolimits_{j=1}^{ N_{\rm L}}\frac{\partial c_{\rm L}^d   D_{\rm L}^{\rm EC}({\bf x})}{\partial x_k}x_j. \nonumber
\end{align}
In addition, according to Lemma \ref{lemma1}, the users with lower delay costs
shall be given lower priority service, in which case the derivative
${\partial D_{\rm H}^{\rm EC}({\bf x})}/{\partial x_k}=0$.
Hence, for the users with higher delay cost and those with
lower delay cost, the AP shall respectively charge them with a
service fee as follows,
\begin{subequations}
\begin{align}
p_{\rm H}=&\sum\limits_{j=1}^{ N_{\rm H}}\frac{\partial c_{\rm H}^d  D_{\rm H}^{\rm EC}({\bf x})}{\partial x_k}x_j +\sum\limits_{j=1}^{ N_{\rm L}}\frac{\partial c_{\rm L}^d   D_{\rm L}^{\rm EC}({\bf x})}{\partial x_k}x_j, \label{eq25a}\\
p_{\rm L}=&\sum\limits_{j=1}^{ N_{\rm L}}\frac{\partial c_{\rm L}^d   D_{\rm L}^{\rm EC}({\bf x})}{\partial x_k}x_j. \label{eq25b}
\end{align}
\end{subequations}

\medskip
\begin{theorem}\label{theorem2}
\textit{For any given price pair $(p_H, p_L)$ satisfying (\ref{eq25a})(\ref{eq25b}) and
the expected delay pair $(D_{{\rm H}}^{\rm EC}, D_{{\rm L}}^{\rm EC})$ satisfying (\ref{eq26a})(\ref{eq26b}),
it is in all the end users' self-interest
to classify their jobs in their correct priority class, i.e.,
\begin{align}
p_{{\rm i}_k}+ c_k^dD_{{\rm i}_k}^{\rm EC}< p_{\bar {\rm i}_k}+ c_k^dD_{\bar{\rm i}_k}^{\rm EC}, {\rm i}_k\in \{{\rm H, L}\}.  \nonumber
\end{align}
where ${\bar {\rm i}_k}={\rm L}$ if ${\rm i}_k={\rm H}$, and ${\bar {\rm i}_k}={\rm H}$ if ${\rm i}_k={\rm L}$.
Moreover, the resulting offloading frequencies maximize the net welfare of all the end users.
}
\end{theorem}
\begin{IEEEproof}
See Appendix B.
\end{IEEEproof}
\medskip

\emph{Remark 2:} Based on Theorem \ref{theorem2}, one can see
that the given price pair $(p_H, p_L)$ is incentive-compatible.
That is, under this pricing mechanism, rational users will choose
to pay the right class service fee and join the right priority class
in terms of social welfare maximization.
\medskip

Substituting the optimal solution derived by Theorem \ref{theorem1}
into (\ref{eq25a})(\ref{eq25b}),
we arrive at the right prices that the AP shall set for the purpose of
inducing users to choose the right priority class.
However, the derivation of the optimal offloading frequency
solution in Theorem \ref{theorem1} requires users' individual profit functions.
In the following subsection, we derive
an evolutional pricing algorithm, which requires no knowledge of individual utility functions
and learns the \textit{correct} prices, such that social welfare is maximized.

\subsection{Learning-based pricing which induces social-optimal offloading}
\begin{figure}[!h]
\centering
\includegraphics[width=3in]{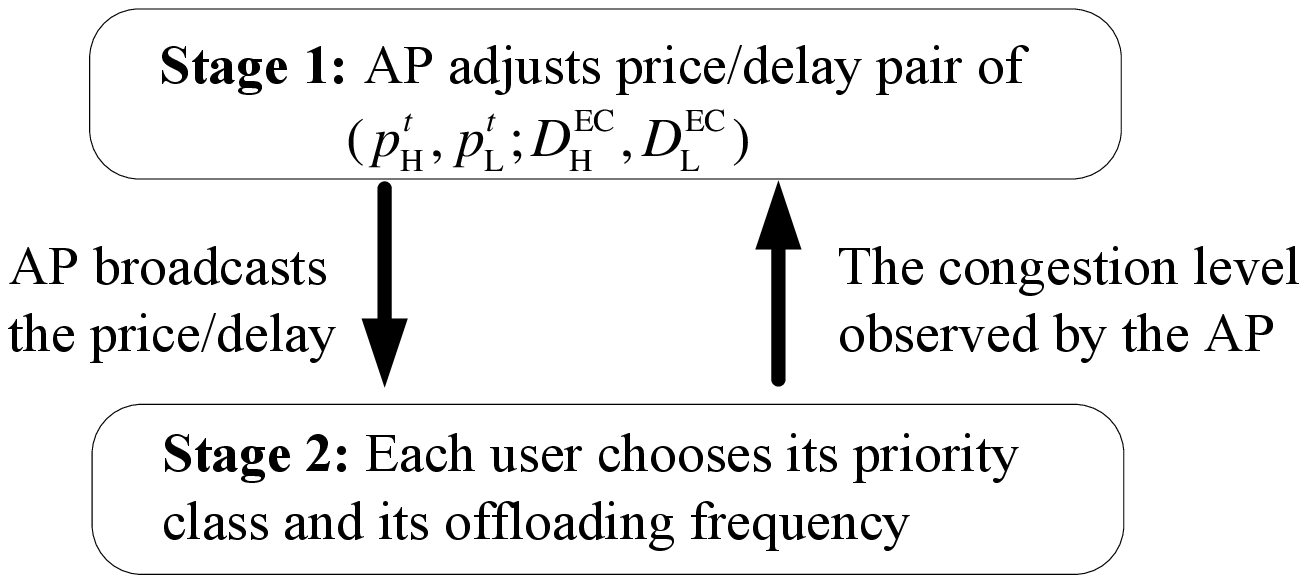}
\DeclareGraphicsExtensions. \caption{Schematic view of the learning-based pricing scheme
for two priority classes of users.}
\label{fig4}
\end{figure}

As shown in Fig. \ref{fig4},
at each time slot $t$, the AP observes $D_{\rm H,EC}^{\rm true,t}$ and $D_{\rm L,EC}^{\rm true,t}$,
the congestion level caused by users' offloading decisions.
Based on the congestion level, the AP computes and
broadcasts the unit service fees $p_{\rm H}^{t}$ and $p_{\rm L}^{t}$,
and delay $D_{\rm H,EC}^t$ and $D_{\rm L,EC}^t$ signals to users.
Viewing those prices and expected delays information, users then
decide their priority class and offloading frequency to maximize
their individual welfare. Please see Fig. \ref{fig4} for the schematic view.

However, it is not easy to extend the evolutional pricing algorithm for the case
with one priority class of users here, since other than inducing users
to offload at the social optimal offloading frequency, the \emph{correct} prices also
need to induce users to choose the \emph{correct} priority class.

The key idea is to adjust prices according to (\ref{eq25a})(\ref{eq25b}).
Some observations are in order. First, given a pair of average delays, i.e.,
$D_{{\rm H}}^{\rm EC}$ and $D_{{\rm L}}^{\rm EC}$, we can respectively determine the
sum offloading rate of the two distinct priority classes of users according
to (\ref{eq26a}) and (\ref{eq26b}), which further decides the corresponding prices
according to (\ref{eq25a})(\ref{eq25b}). Substituting the average delays and prices
into (\ref{eq24}), users can compute the offloading frequency that maximizes their' individual profit.
Second, according to Theorem \ref{theorem2}, given any delay pairs and price pairs
respectively satisfying (\ref{eq26a})(\ref{eq26b}) and (\ref{eq25a})(\ref{eq25b}),
it is in all the end users' self-interest
to classify their jobs in their correct priority class.
That is, those price pairs are incentive-compatible candidates.
Third, the demand function $g(x_k)$
is monotonically deceasing with respect to $x_k$.
As such, if the right hand side of (\ref{eq24}), i.e., the total cost,
is less than that at the social optimal equilibrium, the resulting offloading frequency
$x_k^t> x_k^\star$, which results in a high congestion level.
Based on these observations, we propose Algorithm \ref{alg:NE2} as follows.

\begin{algorithm}
\caption{Learning-based pricing algorithm for two priority classes of users}\label{alg:NE2}
\begin{algorithmic}[1]
\State \textbf{Initialize}: $t \gets 0$, $D_{\rm H,EC}^t \gets \mu_B^{-1}+\varsigma$,
$D_{\rm H,EC}^{{\rm true},t} \gets 0$
\While{$D_{\rm H,EC}^t > D_{\rm H,EC}^{{\rm true},t}$}  \Comment{Learn lower/upper bounds}
\State $D_{\rm H,EC}^{\rm UB} \gets D_{\rm H,EC}^t$
\State $t \gets t+1$, $D_{\rm H,EC}^t \gets \max\{{D_{\rm H,EC}^t}/{2},\mu_B^{-1}\}$
\State $D_{\rm L,EC}^t \gets \mu_BD_{\rm H,EC}^tD_{\rm H,EC}^t$
\State $(p_{\rm H}^{t}, p_{\rm L}^{t}) \gets$ by (\ref{eq26a})(\ref{eq26b}) and (\ref{eq25a})(\ref{eq25b})
\State ${\rm i}_k^\star  \gets {\arg}  \min_{{\rm i}_k} (p_{{\rm i}_k}^t+c_k^d D_{{\rm i}_k,{\rm EC}}^t)$
\State $x_k^{t} \gets g_k^{-1}(p_{{\rm i}_k^\star}^t+c_k^d D_{{\rm i}_k^\star,{\rm EC}}^t)$, $k=1,2,\cdots,N$
\State $D_{\rm L,EC}^{{\rm true}, t} \gets$ by substituting $x_k^{t}$ into (\ref{eq26b})
\While {$D_{\rm L,EC}^t < D_{\rm L,EC}^{{\rm true},t}$}
\State $D_{\rm L,EC}^{\rm LB} \gets D_{\rm L,EC}^t$
\State $t \gets t+1$, $D_{\rm L,EC}^t \gets 2D_{\rm L,EC}^t$, $D_{\rm H,EC}^t \gets D_{\rm H,EC}^t$
\State Execute steps 6-9
\EndWhile
\State \textbf{end}
\State $D_{\rm L,EC}^{\rm UB} \gets D_{\rm L,EC}^t$
\While{$D_{\rm L,EC}^{\rm UB}-D_{\rm L,EC}^{\rm LB} > \epsilon$}  \Comment{Bisection search}
\State $t \gets t+1$,
$D_{\rm L,EC}^t \gets ({D_{\rm L,EC}^{\rm UB}+D_{\rm L,EC}^{\rm LB}})/{2}$, $D_{\rm H,EC}^t \gets D_{\rm H,EC}^t$
\State Execute steps 6-9
\If {$D_{\rm L,EC}^t < D_{\rm L,EC}^{{\rm true},t}$}
$D_{\rm L,EC}^{\rm LB} \gets D_{\rm L,EC}^t$
\Else \ $D_{\rm L,EC}^{\rm UB} \gets D_{\rm L,EC}^t$
\EndIf
\EndWhile\label{euclidendwhile}
\State \textbf{end}                 \Comment{The stop threshold $\epsilon = 0.01$}
\State $D_{\rm H,EC}^{{\rm true}, t} \gets$ by substituting $x_k^{t}$ into (\ref{eq26a})
\EndWhile
\State \textbf{end}
\State $D_{\rm H,EC}^{\rm LB} \gets D_{\rm H,EC}^t$
\While{$D_{\rm H,EC}^{\rm UB}-D_{\rm H,EC}^{\rm LB} > \epsilon$}  \Comment{Bisection search}
\State $t \gets t+1$, $D_{\rm H,EC}^t \gets (D_{\rm H,EC}^{\rm UB}+D_{\rm H,EC}^{\rm LB})/2$
\State Execute steps 5-22
\If {$D_{\rm H,EC}^t < D_{\rm H,EC}^{{\rm true},t}$}
$D_{\rm H,EC}^{\rm LB} \gets D_{\rm H,EC}^t$
\Else \ $D_{\rm H,EC}^{\rm UB} \gets D_{\rm H,EC}^t$
\EndIf
\EndWhile
\State \textbf{end}
\State \textbf{return} $p_{\rm H} \gets p_{\rm H}^t$, $p_{\rm L} \gets p_{\rm L}^t$,
$D_{\rm H}^{\rm EC} \gets D_{\rm H, EC}^t $,
$D_{\rm L}^{\rm EC} \gets D_{\rm L, EC}^t $ 
\end{algorithmic}
\end{algorithm}

Specifically, the whole algorithm consists of two embedded loops, where
in the outer loop the AP updates the expected average delay of $D_{\rm H, EC}^t$ for
the high priority queue,
while in the inner loop the AP updates the expected average delay of $D_{\rm L, EC}^t$
for the low priority queue.

In the inner loop and at each time slot,
the AP observes $D_{\rm L,EC}^{\rm true,t-1}$, the congestion
level of the low priority queue, caused by users' offloading decisions.
Based on the observed congestion level, the AP computes
the expected average delay $D_{\rm L, EC}^t$ for the low priority queue, which, combined with
$D_{\rm H, EC}^t$, decides the unit service fee $p_{\rm H}^t$ and $p_{\rm L}^t$.
Following which, the AP broadcasts the service fee and expected delay signals to users.
Viewing the signals, users decide which priority class to join and determine
their offloading frequency $x_k^t$ based on (\ref{eq24})
to maximize their individual welfare.
The above process iterates until convergence when
the resulting congestion level $D_{\rm L,EC}^{\rm true,t}$ equals
the expected average delay set, i.e., $D_{\rm L, EC}^t$.

In the outer loop and at each iteration,
the AP observes $D_{\rm H,EC}^{\rm true,t-1}$, the congestion
level of the high priority queue. Based on the observed congestion level, the AP computes
the expected average delay $D_{\rm H, EC}^t$ for the high priority queue.
The inner loop (steps 5-26) is then executed until the congestion level of
the low priority queue converges to its expected average delay.
Back to outer loop, the AP will check if $D_{\rm H,EC}^{\rm true,t}$, the observed congestion level of the
high priority queue has converged to $D_{\rm H, EC}^t$, its expected average delay.
If not, the outer loop will change its expected average delay again, and
continue with the inner loop (steps 5-26). Otherwise, the entire process
stops and we have arrived at the optimal prices and expected delays of the two priority classes
that maximize the system-wide welfare.

The following theorem proves that Algorithm \ref{alg:NE2} converges.

\medskip
\begin{theorem}\label{theorem3}
\textit{Algorithm \ref{alg:NE2} finally converges to a socially optimal point where
users classify their jobs in their correct priority class and offload at
the offloading frequencies which jointly maximize the system-wide welfare.
}
\end{theorem}
\begin{IEEEproof}
See Appendix C.
\end{IEEEproof}
\medskip

\section{Numerical Results}
In this section, we provide numerical simulations which substantiate our theoretical results. The simulations show the convergence and optimality of our proposed pricing mechanism, and give us some insight into the system performance at the optimal allocation.

A number of 100 end users are uniformly placed at random
on a ring of radius $10 \le d_k \le 75$ (unit: meters) whose center is located at the AP.
The path loss exponent is $\alpha=3.5$.
The channels are assumed to be identically distributed (i.i.d.) and $|h_k|^2 \sim {\rm exp}(1)$.
The heterogeneous scenario is considered, where users have varying computational delay and energy requirements,
and a priority based pricing mechanism offering two priority classes (Algorithm \ref{alg:NE2} in Section 4.2) is proposed to maximize the overall welfare of heterogeneous users.
Without loss of generality, we assume that the first 50 users
set their computational weights as $c_k^d=0.9$ and the
other 50 users set their weights as $c_k^d=0.1$.
The key parameters are summarized in Table \ref{tab:table2}.

\begin{table}[h!]
\centering
\caption{Simulation Parameters \cite{Jeongho15,ChenFeng17,Sokol12}}
\medskip
\label{tab:table2}
    \begin{tabular}{||c|c || c|c ||}
 \hline
 \textbf{Parameter} & \textbf{Value} &\textbf{Parameter} & \textbf{Value} \\ [0.5ex]
 \hline\hline
 $W$ & 360KHz & $\kappa_m$ & $10^{-27} \rm Watt\cdot {s}^3/{cycles}^3$\\
 $P_{\rm tr}$ & 100mW & $B_a$ & $8250\ \rm cycles/bit$\\
 $\sigma^2$ & -40dBm & $L_a$ & 100K Bytes\\
 $\alpha$ & 3.5 & $\lambda_a$ & $0.01\ \rm jobs/s$\\
 $f_B$ & 3GHz & $c_k^d$ & \{0.9,0.1\}\\
 $f_m$ & 0.5GHz & $c_k^e$ & \{0.1,0.9\} \\[1ex]
 \hline
 \end{tabular}
\end{table}

\begin{figure}[!t]
\centering
\includegraphics[width=3in]{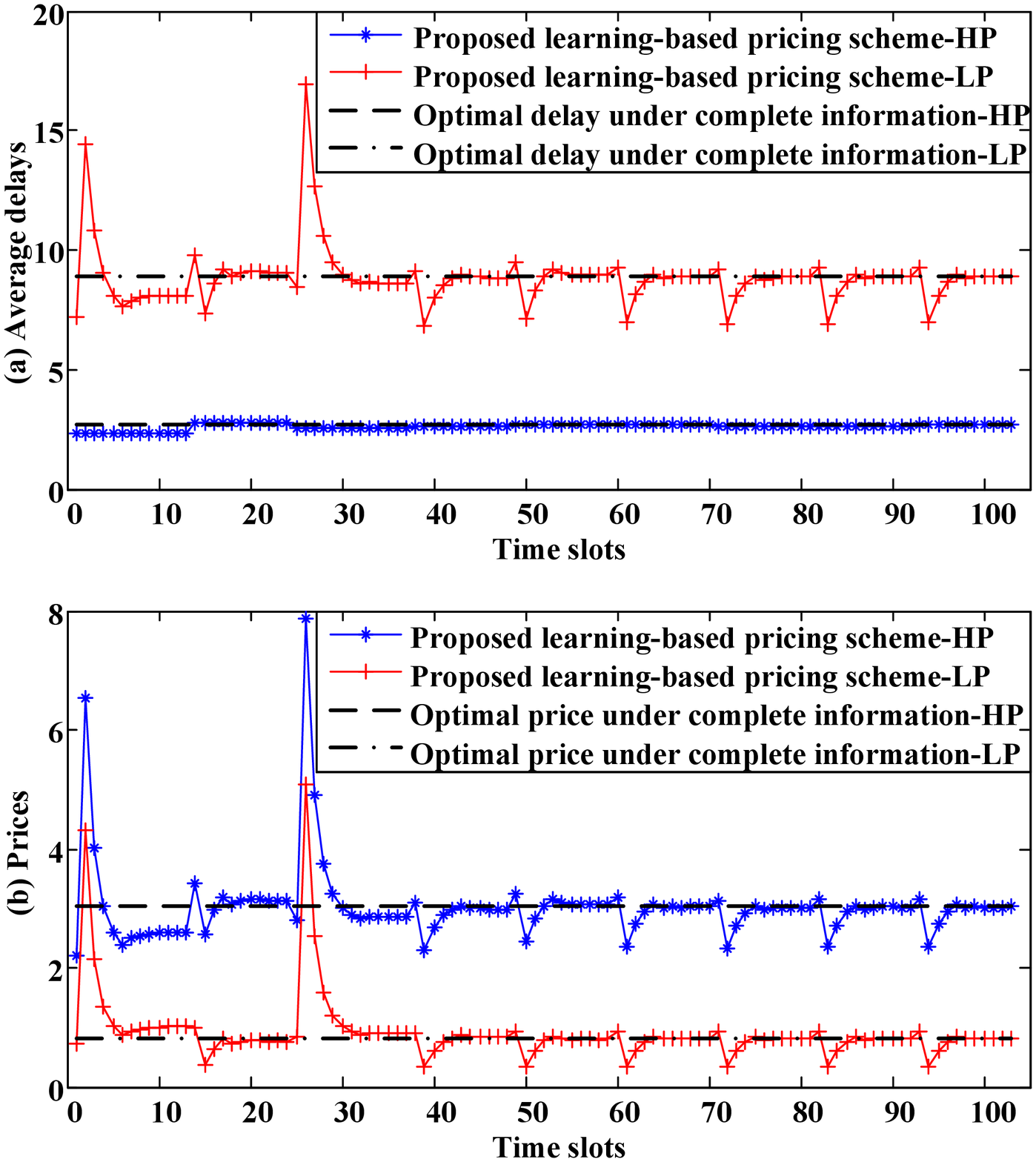}
\DeclareGraphicsExtensions. \caption{(a) Posted delay v.s. time; (b) Posted price v.s. time}
\label{fig7}
\vspace* {-6pt}
\end{figure}
In Fig. \ref{fig7} we plot how the average delays and prices for both the low and high priority classes varying over time. Recall that Algorithm \ref{alg:NE2} consists of two embedded loops. In the inner loop the expected average delay of the low priority class $D^t_{\rm L,EC}$ is updated until the resulting congestion level $D^{{\rm true},t}_{\rm L,EC} $ equals $D^t_{\rm L,EC}$. While in the outer loop, the expected average delay of the high priority class $D^t_{\rm H,EC}$ is updated, until the resulting congestion level $D^{{\rm true},t}_{\rm  H,EC} $ equals $D^t_{\rm H,EC}$. This embedded loop structure can be visualised by the algorithm's trajectory in the figures. From Fig. \ref{fig7}, we can see that our mechanism learns the optimal delays and prices for both classes. The AP learns these optimal values without knowledge of users' utility functions (which encapsulate private information like battery states). It can be seen that the optimal price for the high priority class is around 3 times higher, because i) they are paying for the higher delay they cause to the low priority users (see eq. (\ref{eq25a}) and (\ref{eq25b})),
ii) they have low delay tolerance and would be willing to pay a higher price for a lower average delay.

\begin{figure}[!t]
\centering
\includegraphics[width=3in]{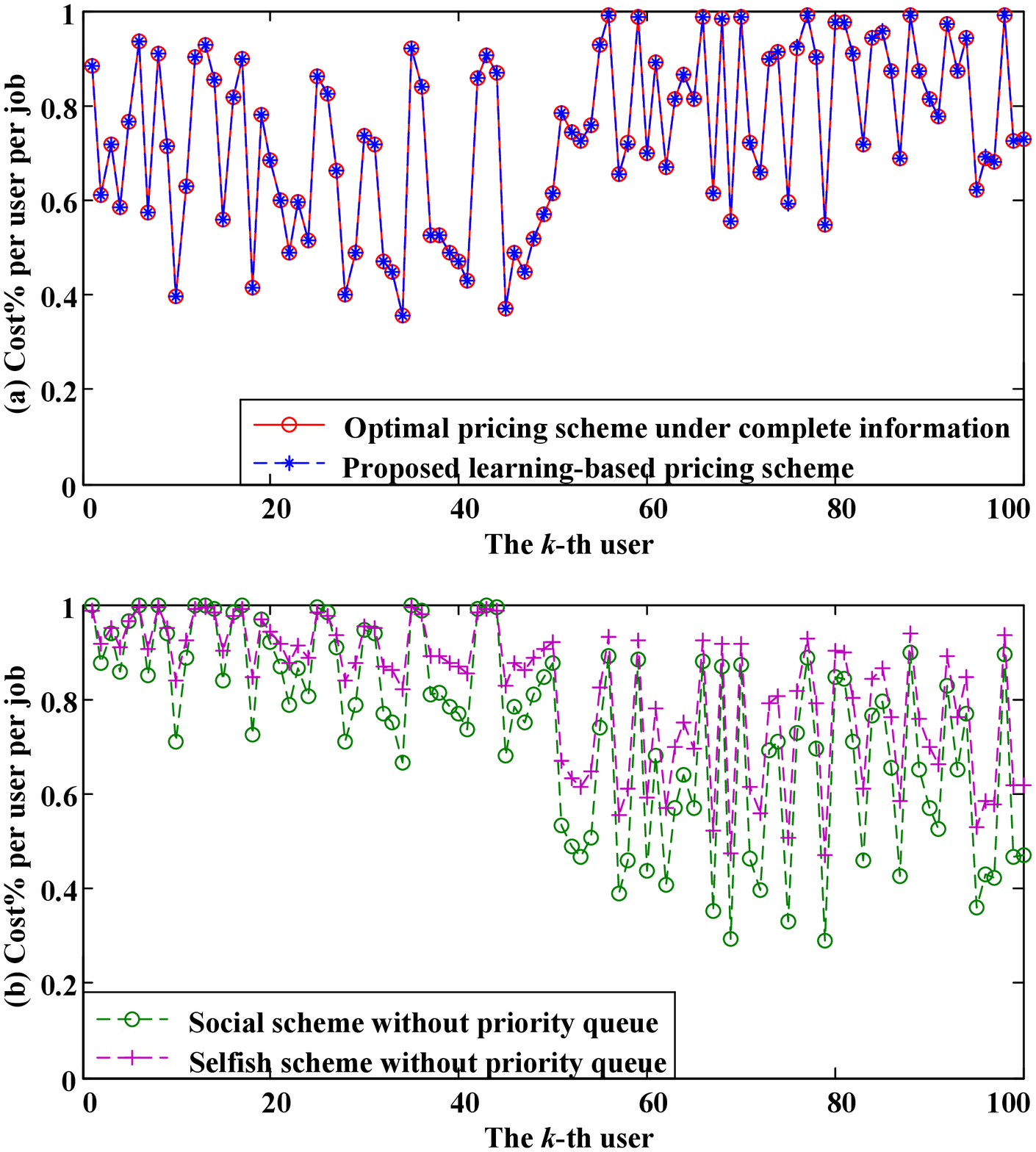}
\DeclareGraphicsExtensions. \caption{Job execution costs ($\%$) via different schemes.}
\label{fig8}
\vspace* {-6pt}
\end{figure}

Fig. \ref{fig8} illustrates the individual users' cost per job as a percentage of that by local computing only. The subplot Fig. \ref{fig8}(a) illustrates that at the equilibrium point of our mechanism, the users achieve the optimal cost savings. As comparison, in Fig. \ref{fig8}(b) we also plot the individual users' cost by
other schemes without priority queue, the social scheme and the selfish scheme.
Specifically, in the social scheme we maximize the net welfare of users,
while in the selfish problem users maximize their individual interest.
Similar to Subsection 4.1, optimal solutions of these two schemes could be respectively obtained
by solving the equations arising from the derivatives of the net welfare and users' individual interest.
From Fig. \ref{fig8}, we can see that as compared with the proposed scheme, under the schemes without priority
queue, users with stricter delay requirement (i.e., bigger computational weight) suffers from profit loss.
This can be explained as follows. Given a same expected value of average delay,
users with a bigger computational weight pay more due to the higher delay cost. As such,
their demand in offloading decreases. In contrast, users with a smaller computational weight
are more willing to offload. Therefore, the former group of users cannot enjoy well the
cost saving brought by edge computing.

%

\begin{figure}[!t]
\centering
\includegraphics[width=3in]{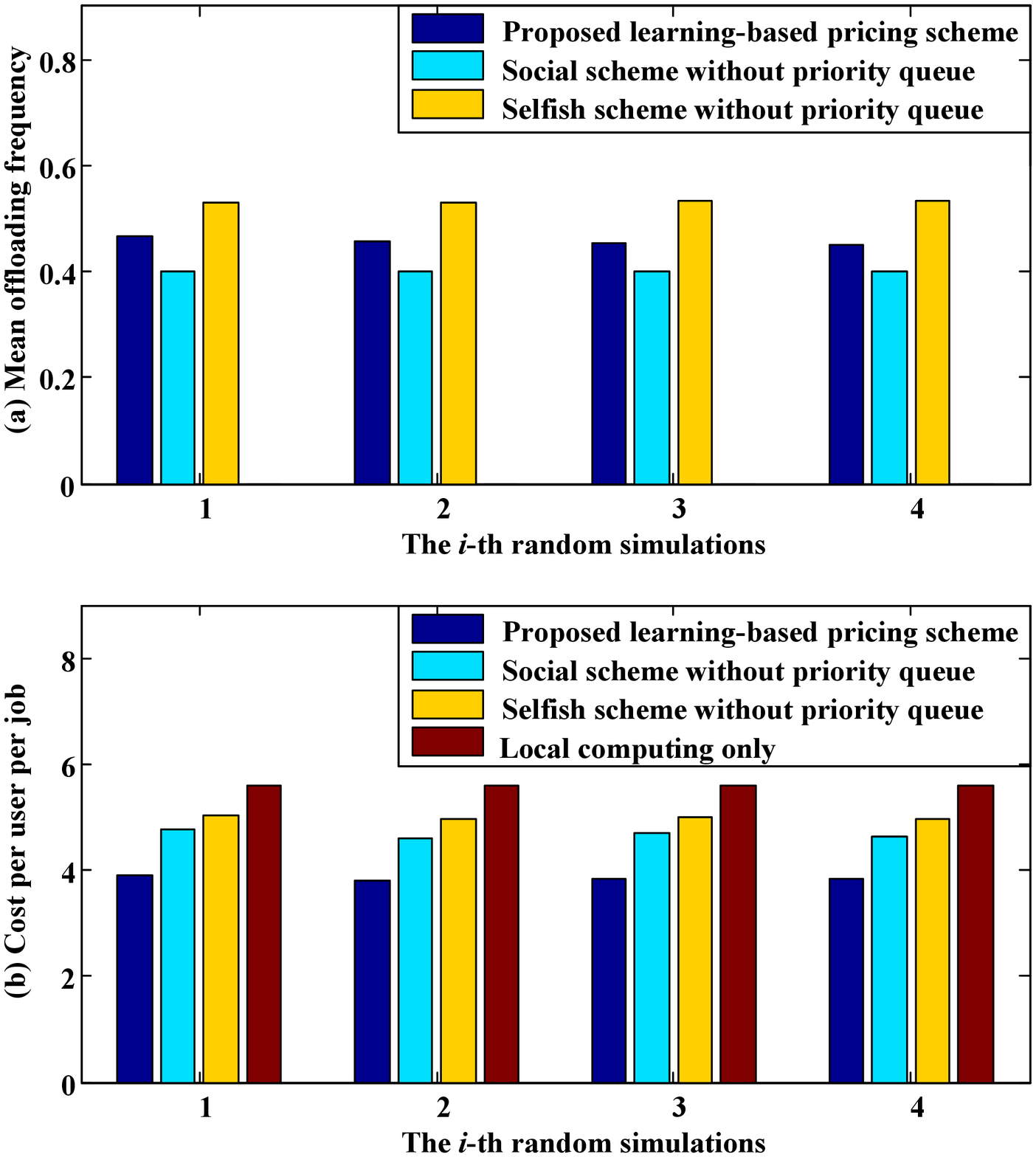}
\DeclareGraphicsExtensions. \caption{(a) Mean offloading frequency; (b) Average cost.}
\label{fig10}
\vspace* {-6pt}
\end{figure}

We conclude this section with Fig. \ref{fig10}, in which
we run the simulation for the four different schemes. In each simulation, we place
end users uniformly, and respectively plot the resulting mean offloading frequency of users as well as
the job execution cost of the different schemes.
From Fig. \ref{fig10}, we can see that by providing the option of priority,
users offload more often, which results in bigger cost savings (almost reduced $30\%$ of the cost, as
compared to local computing only).
In contrast, users offload more under the selfish scheme, as compared to the social scheme,
however, they suffer from an improper high congestion level. As such, they suffer welfare loss on average.

\section{Conclusion}
In this paper, we have proposed an incentive-aware offloading
mechanism for an MEC system, where an
(AP) of finite computing power serves
two classes of resource-hungry and selfish users.
As users have heterogeneous service requirements,
our mechanism provides the option of priority, serving the high
priority users first for a higher price.
We have characterized in semi-closed form the socially
optimal prices, and proposed a learning-based pricing
mechanism for the incomplete information scenario.
Further, we have proved
that the proposed mechanism converges to the optimal prices
and expected edge delays, such that at equilibrium,
self-interested users choose the correct priority class and
make socially optimal offloading decisions.

In this work we have assumed that the MEC server executes jobs one
after another.
Future work will take parallel edge computing into account.
Also, we have considered the static scenario where end users (end devices) do not move.
It would be interesting to use reinforcement learning to allow the AP to learn the behaviors of end users
such as the mobility pattern and the job arrival pattern, so that it could make dynamic pricing
decisions based on the states of a dynamic environment, and we plan to tackle this in our future work.

\appendices
\section{Proof of \emph{Theorem 1}} \label{appA}
According to Lemma 1, for the purpose of maximizing the net
welfare of users, a user with stricter delay requirement shall choose the higher priority service;
otherwise, it shall choose the lower priority service.
As such, the edge computing delay can be rewritten as follows,
\begin{align}
D_{{\rm i}_j}^{\rm EC}({\bf x}) = \left\{ {\begin{array}{*{20}{c}}
{D_{\rm H}^{\rm EC}({\bf x}),\ {\rm if} \ c_j^d = c_{\rm{H}}^d,}\\
{D_{\rm L}^{\rm EC}({\bf x}),\ {\rm if} \ c_j^d = c_{\rm{L}}^d,}
\end{array}} \right.  \label{eqA1}
\end{align}
where ${D_{\rm H}^{\rm EC}}({\bf x})$ and ${D_{\rm L}^{\rm EC}}({\bf x})$ are given by (17a) and (17b), respectively. Substituting (\ref{eqA1}) into (18) yields
(19).

Besides, taking the derivative of the formula in (17a) with respect to $x_k$ yields
\begin{align}
\dfrac{{\partial  D_{\rm{H}}^{{\rm{EC}}}({\bf{x}})}}{{\partial {x_k}}} = \left\{ {\begin{array}{*{20}{c}}
{{{{\lambda _a}}}/{\Psi_{\rm H}^2}}, &{\rm if} \ c_k^d = c_{\rm{H}}^d,\\
0, &{\rm if} \ c_k^d = c_{\rm{L}}^d.
\end{array}} \right. \label{eqA2}
\end{align}
Similarly, taking the derivative of the formula in (21b) yields
\begin{align}
\dfrac{{\partial D_{\rm{L}}^{{\rm{EC}}}({\bf{x}})}}{{\partial {x_k}}} = \left\{ {\begin{array}{*{20}{c}}
{\dfrac{{{\lambda _a}{\mu _B}({\Psi _{\rm{H}}} + \Psi )}}{{{\Psi ^2}\Psi _{\rm{H}}^2}},\;}&{{\rm{if}}\;c_k^d = c_{\rm{H}}^d,}\\
{\dfrac{{{\lambda _a\mu _B}}}{{{\Psi ^2}\Psi _{\rm{H}}}},\;}&{{\rm{if}}\;c_k^d = c_{\rm{L}}^d.}
\end{array}} \right. \label{eqA3}
\end{align}
Substituting (\ref{eqA2}) and (\ref{eqA3}) into (19), we have (20).
This completes the proof.

\section{Proof of \emph{Theorem 2}} \label{appB}
In this section, we will prove the following two equations, thus completing the proof of Theorem 2.
\begin{subequations}
\begin{align}
p_{\rm H}+ c_{\rm H}^dD_{\rm H}^{\rm EC}< p_{\rm L}+ c_{\rm H}^dD_{\rm L}^{\rm EC}, \label{eqB1a}\\
p_{\rm L}+ c_{\rm L}^dD_{\rm L}^{\rm EC}< p_{\rm H}+ c_{\rm L}^dD_{\rm H}^{\rm EC}. \label{eqB1b}
\end{align}
\end{subequations}

Combining (22a)(22b) and (\ref{eqA2})(\ref{eqA3}), we arrive at
\begin{subequations}
\begin{align}
p_{\rm H}-p_{\rm L}=&\sum\nolimits_{j = 1}^{{N_{\rm{H}}}} {\dfrac{{{\lambda _a}c_{\rm{H}}^d{x_j}}}{{\Psi _{\rm{H}}^2}}}  + \sum\nolimits_{j = 1}^{{N_{\rm{L}}}} {\dfrac{{{\lambda _a\mu _B}c_{\rm{L}}^d{x_j}({\Psi _{\rm{H}}} + \Psi )}}{{{\Psi ^2}\Psi _{\rm{H}}^2}}} \nonumber \\
&-\sum\nolimits_{j = 1}^{{N_{\rm{L}}}} {\dfrac{{{\lambda _a}{\mu _B}c_{\rm{L}}^d{x_j}}}{{{\Psi ^2}{\Psi _{\rm{H}}}}}}\nonumber \\
=&{c_{\rm{H}}^d\frac{{\sum\nolimits_{j = 1}^{{N_{\rm{H}}}} {{\lambda _a}{x_j}} }}{{\Psi _{\rm{H}}^2}} + \frac{{{\mu _B}c_{\rm{L}}^d}}{{{\Psi ^2}{\Psi _{\rm{H}}}}}\frac{{\Psi ({\Psi _{\rm{H}}} - \Psi )}}{{{\Psi _{\rm{H}}}}}} \label{eqB2a} \\
< & {\frac{{{\mu _B}c_{\rm{H}}^d}}{{\Psi {\Psi _{\rm{H}}}}} - \frac{{c_{\rm{H}}^d}}{{{\Psi _{\rm{H}}}}}{\rm{ = }}c_{\rm{H}}^d(D_{\rm{L}}^{{\rm{EC}}} - D_{\rm{H}}^{{\rm{EC}}})}, \label{eqB2c}
\end{align}
\end{subequations}
where (\ref{eqB2a}) comes from the fact that
${\Psi _{\rm{H}}} - \Psi=\sum\nolimits_{j = 1}^{{N_{\rm{L}}}} {{\lambda _a}{x_j}} $;
the inequality in (\ref{eqB2c}) comes from the fact that $c_{\rm{H}}^d>c_{\rm{L}}^d$.
Similarly, it can be verified that the last inequality in (\ref{eqB2c})
can be replaced by $p_{\rm H}-p_{\rm L}> c_{\rm{L}}^d(D_{\rm{L}}^{{\rm{EC}}} - D_{\rm{H}}^{{\rm{EC}}})$. This completes the proof.

\section{Proof of \emph{Theorem 3}} \label{appC}
Let $(D_{\rm L,EC}^\star, D_{\rm H,EC}^\star)$ and $(p_{\rm L}^\star, p_{\rm H}^\star)$
respectively represent the optimal delays and prices arising from
the system-wide welfare offloading, denoted by $x_k^\star$.
In the following, we will respectively show the convergence
of the inner and the outer loop, followed by the proof that Algorithm 1
converges to the offloading decisions that jointly maximzie the system-wide welfare.

\emph{(1) For any given fixed average delay of the high priority queue, i.e., $D_{\rm H,EC}^t=D_{\rm H,EC}$,
the inner loop converges to a point where $D_{\rm L,EC}^{{\rm true},t}=D_{\rm L,EC}^t$.}
Some observations are in order. First, when setting $D_{\rm L,EC}^t = D_{\rm H,EC}+\varsigma$
we get $D_{\rm L,EC}^{{\rm true},t}>D_{\rm L,EC}^t$, and
when setting $D_{\rm L,EC}^t =\infty-\varsigma$ we get $D_{\rm L,EC}^{{\rm true},t}<D_{\rm L,EC}^t$.
Here $\varsigma$ is an almost zero but positive value.
Second, $D_{\rm L,EC}^{{\rm true},t}$ is a decreasing function of
$D_{\rm L,EC}^t \in (D_{\rm H,EC}, \infty)$.
This can be proved as follows.
Assume that $D_{\rm L,EC}^{t_1}<D_{\rm L,EC}^{t_2}$.
Then, $p_{\rm H}^{t_1} < p_{\rm H}^{t_2}$ and $p_{\rm L}^{t_1} < p_{\rm L}^{t_2}$.
This, combined with the fact that the demand function $g_k(x_k)$ in (21)
is monotonically decreasing, indicates that
$x_k^{t_1}>x_k^{t_2}$. Hence, it holds that $D_{\rm L,EC}^{{\rm true},{t_1}} > D_{\rm L,EC}^{{\rm true},{t_2}}$.
Concluding the above observations, the inner loop converges to a point where $D_{\rm L,EC}^{{\rm true},t}=D_{\rm L,EC}^t$, where, for the ease of exposition, we respectively
use $D_{\rm L,EC}^e$, $p_{\rm L}^e$ and $p_{\rm H}^e$ to
represent the equilibrium delay and prices.

\emph{(2) The outer loop converges to the optimal point where $D_{\rm H,EC}^{{\rm true},t}=D_{\rm H,EC}^t=D_{\rm H,EC}^\star$.}
In the following, we will prove by contradiction that $D_{\rm H,EC}^{{\rm true},t}<D_{\rm H,EC}^t$ holds true only if $D_{\rm H,EC}^t>D_{\rm H,EC}^\star$; the argument for the inverse case is similar.
Concluding, the outer loop converges to a point where $D_{\rm H,EC}^{{\rm true},t}=D_{\rm H,EC}^t=D_{\rm H,EC}^\star$.

Assume $D_{\rm H,EC}^t \le D_{\rm H,EC}^\star$. Then, we have the following
inference.

i) Its embedded inner loop converges to a point where
\begin{align}
p_{\rm L}^e+ c_{\rm L}^dD_{\rm L, EC}^e \le p_{\rm L}^\star+ c_{\rm L}^dD_{\rm L, EC}^\star; \label{eqC1}
\end{align}
otherwise, assume that $p_{\rm L}^e+ c_{\rm L}^dD_{\rm L, EC}^e > p_{\rm L}^\star+ c_{\rm L}^dD_{\rm L, EC}^\star$, and since the demand function $g_k(x_k)$ in (21)
is monotonically decreasing with respect to $x_k$, the resulting offloading frequency of users
associated with the low priority queue satisfies $x_k < x_k^\star$.
This, combined with the assumption $D_{\rm H,EC}^t \le D_{\rm H,EC}^\star$, indicates that
$p_{\rm L}^e+ c_{\rm L}^dD_{\rm L, EC}^e < p_{\rm L}^\star+ c_{\rm L}^dD_{\rm L, EC}^\star$, which, however,
contradicts with the assumption.

ii) It holds that $D_{\rm H,EC}^{{\rm true},t} \ge D_{\rm H,EC}^t $.
This can be explained as follows. By Algorithm 1, the posted average edge delay and price
candidates always respectively satisfying (17a)(17b) and (22a)(22b),
based on which it can be verified that
\begin{subequations}
\begin{align}
& p_{\rm H}^e+c_{\rm H}^dD_{\rm H,EC}^t \nonumber\\
&={(c_{\rm H}^d-c_{\rm L}^d)\mu_B}{{D_{\rm H,EC}^t}^2}+p_{\rm L}^e+ c_{\rm L}^dD_{\rm L, EC}^e  \label{eqC2a}\\
& \le {(c_{\rm H}^d-c_{\rm L}^d)\mu_B}{{D_{\rm H,EC}^\star}^2}+p_{\rm L}^\star+ c_{\rm L}^dD_{\rm L, EC}^\star \label{eqC2b}\\
& = p_{\rm H}^\star+c_{\rm H}^d D_{\rm H,EC}^\star, \label{eqC2d}
\end{align}
\end{subequations}
where (\ref{eqC2b}) comes from the inference (\ref{eqC1}) and the assumption $D_{\rm H,EC}^t \le D_{\rm H,EC}^\star$.
In addition, since the demand function $g_k(x_k)$ is monotonically decreasing with respect to $x_k$, the resulting offloading frequency of users associated with the high priority queue satisfies $x_k \ge x_k^\star$.
As such, it holds that $D_{\rm H,EC}^{{\rm true},t} \ge D_{\rm H,EC}^\star \ge D_{\rm H,EC}^t $.

The second inference contradicts with the fact that $D_{\rm H,EC}^{{\rm true},t} < D_{\rm H,EC}^t $.
This completes the proof that $D_{\rm H,EC}^{{\rm true},t}<D_{\rm H,EC}^t$ holds true only if $D_{\rm H,EC}^t>D_{\rm H,EC}^\star$.

\emph{(3) Combining the above analysis (1) and (2), one can see that Algorithm 1
converges to a point where $D_{\rm L,EC}^{{\rm true},t}=D_{\rm L,EC}^t$ and $D_{\rm H,EC}^{{\rm true},t}=D_{\rm H,EC}^t$. This, combined with Theorem 1, indicates that Algorithm 1
converges to offloading decisions that jointly maximize the system-wide welfare.}
This completes the proof.

\bibliography{mybib}
\bibliographystyle{IEEEtran}

\end{document}